\documentclass[12pt]{article}
\usepackage{amssymb}

\usepackage{makeidx}
\usepackage{jeep}
\usepackage{mcite}
\usepackage{graphics}
\usepackage[dvips]{graphicx}

\input tcilatex
\makeindex

\begin{document}

%
\date{\ }
\title{ENTROPY: MYSTERY AND CONTROVERSY \\
Plethora of Informational-Entropies and \\
Unconventional Statistics}

\author{ Roberto Luzzi, \'{A}urea R. Vasconcellos, and J. Galv\~{a}o Ramos \\
\ \\
{\small \emph{Instituto de F\'{\i}sica `Gleb Wataghin',}}\\
{\small \emph{Universidade Estadual de Campinas, Unicamp}}\\
{\small \emph{13083-970 Campinas, S\~ao Paulo, Brazil}}}
\maketitle
%

\begin{quotation}
%
%
\thispagestyle{empty}
\noindent

Some general considerations on the notion of entropy in physics are presented. An
attempt is made to clarify the question of the differentiation between physical entropy
(the Clausius-Boltzmann one) and quantities called entropies associated to Information
Theory, which are in fact generating functionals for the derivation of probability
distributions and not thermodynamic functions of state. The role of them in the
construction of the so-called Unconventional Statistical Mechanics, and the meaning
and use of the latter, is discussed. Particular attention is given to the situation involving
far-from-equilibrium systems.

%
\end{quotation}

\newpage

\section{INTRODUCTION: On Physical Entropy vs.\protect\linebreak
Informational Entropies}

\textit{Entropy -- The Curse of Statistical Mechanics?}, wrote Herman Haken
at some point in his book on Synergetics \cite{Hak78}. G. N. Alekseev
commented \ \cite{Ale86} that a popular-science author of the early
twentieth century wrote that ``Entropy is the name of the Queen's shadow.
Face to face with this phenomenon, Man cannot help feeling some vague fear:
Entropy, like an evil spirit, tries to diminish or annihilate the best
creations of that gracious spirit, Energy. We are all under the protection
of Energy and all potential victims of the latent poison of Entropy ...''.
On the other hand, we can recall the sentence of the great Ludwig Boltzmann
that ``Thus, the general struggle for life is neither a fight for basic
material ... nor for energy ... but for entropy becoming available by the
transition from the hot sun to the cold earth'' \cite{Mai94}.

Moreover, Louis de Broglie \cite{Bro79} expressed that ``En Thermodinamique
classique, on introduit pour \'{e}noncer le second principle de cette
science la grandeur `Entropie' dont la signification physique restait si
obscure que Henry Poincar\'{e} la qualifiait de `prodigeusement abstraite'
''. As expressed by de Broglie entropy is related to the second law of
Thermodynamics, and on that the French philosopher Henri Bergson \cite
{Ber907} called this second law (or Entropy law) the most ``metaphysical''
of all the laws of Nature. Georgescu-Roegen \cite{GR71} commented that from
the epistemological point of view, the Entropy Law may be regarded as the
greatest transformation ever suffered in physics, and that it is the basis
of the economy of life at all levels. Also, that nobody is able to live
without the sensation of the flux of entropy, that is, of that sensation
that under diverse forms regulate the activities related to the maintenance
of the organism. In the case of mammals this includes not only the
sensations of cold and warm, but, also, the pangs of hunger or the
satisfaction after a meal, the sensations of being tired or rested, and many
others of the same type.

Returning to de Broglie \cite{Bro79}, he wrote that ``C'est Boltzmann qui,
en d\'{e}velopant les id\'{e}es de la Thermodinamique statistique, nos a
donn\'{e} le v\'{e}ritable sens de cette grandeur en montrant que l'entropie
$S$ de l'etat d'un corps est reli\'{e}e \`{a} la probabilit\'{e} $P$ de cet
etat par la c\'{e}l\`{e}bre formule: $S=k_{B}\ lnP$'', and that ``en
M\'{e}canique analytique [...] la v\'{e}ritable signification do principle
de Hamilton est la suivante: Le mouvement d'un corps est celui que
poss\`{e}de la plus grande probabilit\'{e} thermodynamique dans les
conditions auxquelles is est soumis. Je pense que cette conception de la
nature profunde du principle de Hamilton jette une flot de lumi\`{e}re sur
sons v\'{e}ritable sens, analogue \`{a} celui que jette la formule de
Boltzmann sur la signification de l'entropie''.

The law of increase of entropy addresses the question of irreversibility of
processes in nature, that is, the famous ``time-arrow'' according to which
all macroscopic events have a preferred time direction. The question has
recently been addressed in two relevant papers in Physics Today. On the one
hand, Joel Lebowitz \cite{Leb93} expressed that ``Boltzmann's thoughts on
this question have withstood the test of time. Given the success of Ludwig
Boltzmann's statistical approach in explaining the observed irreversible
behavior of macroscopic systems in a manner consistent with their reversible
microscopic dynamics, it is quite surprising that there is still so much
confusion about the problem of irreversibility''. In the other article, by
Elliot Lieb and Jacob Yngvason \cite{LY93}, it is raised the point that
``the existence of entropy, and its increase, can be understood without
reference to either statistical mechanics or heat engines''.

Points of view associated to the latter one go back in time, and it can be
mentioned the approach of Elias Gyftopoulos et al. \cite{GC97,Gyf02}: The
basic idea seems to consist in that, instead of regarding mechanics and
thermodynamics as two different theories, belonging to two different levels
of fundamentality, can be considered a new hypothesis, namely, that
mechanics and thermodynamics are two particular aspects of a more general
fundamental theory \cite{HG76}. On entropy it has been noticed \cite{GC97}
that ``In his extensive and authoritative review, Wehrl \cite{Weh78} writes
`It is paradoxical that although entropy is one of the most important
quantities in physics, its main properties are rarely listed in the usual
textbooks on statistical mechanics'... The lack of specificity has resulted
in a \textit{plethora of expressions purporting to represent the entropy of
thermodynamics} (emphasis is ours), and perhaps influenced von Newmann to
respond to Shannon's question `what should I call $-\sum\limits_{i}p_{i}\ln
p_{i}$?' by saying `You should call it `entropy' for two reasons: first, the
function is already in use in thermodynamics under that name; second, and
more importantly, most people don't know what entropy really is, and if you
use that word entropy you will win every time' ''. Moreover, \cite{Gyfpc}
that quantum theory admits not only probability distributions described by
wave functions or projectors, but also probability distributions that are
described \ by \textit{density operators }$\varrho $\textit{\ which are not
statistical averages of projectors}. Said differently, $\varrho $ can be
represented only by a \textit{homogeneous ensemble}, namely, an ensemble
every member of which is characterized by the same $\varrho $ as the whole
ensemble. For projectors (wave functions) the concept of a homogeneous
ensemble was conceived by von Neumann. This discovery plus all the
requirements that the entropy of thermodynamics must satisfy both for
reversible and irreversible processes yield that the only expression for
entropy is $S=-k_{B}Tr\left\{ \varrho \ln \varrho \right\} $, provided that $%
\varrho $ is representable by a homogeneous ensemble. In addition, the
discovery requires a new complete equation of motion different from the
Schr\"{o}dinger equation. Such an equation has been conceived by Beretta
\cite{Ber84}. In both, an unified quantum theory of mechanics and
thermodynamics, and in a novel exposition of thermodynamics without
reference to quantum mechanics, it is proved that entropy is an intrinsic
(inherent) property of each constituent of any system (both macroscopic and
microscopic), in any state (both thermodynamic equilibrium and not
thermodynamic equilibrium) in the sense that inertial mass is an intrinsic
property of each constituent of any system in any state \cite{GB91}. We also
noticed that, Edwin Jaynes showed that the expression above can be
identified with Clausius' entropy \cite{Jay65}, and with Boltzmann-Planck
expression for isolated systems in equilibrium, namely, the celebrated $%
S=k_{B}\ lnW$.

On this, in a recent journalistic article by A. Cho in \textit{Science} \cite
{Cho02} we can read that ``Near the middle of Vienna's sprawling Central
Cemetery stands the imposing tomb of Ludwig Boltzmann, the $19$th century
Austrian physicist who first connected the motions of atoms and molecules to
temperature, pressure, and other properties of macroscopic objects. Carved
in the gravestone, a single short equation serves as the great man's
epitaph: $S=k_{B}\ lnW$. No less important than Einstein's $E=mc^{2}$, the
equation provides the mathematical definition of entropy, a measure of
disorder that every physical system strives to maximize. The equation serves
as the cornerstone of ``statistical mechanics'', and it has helped
scientists decipher phenomena ranging from the various states of matter to
the behavior of black holes to the chemistry of life''. In continuation A.
Cho comments on the existence of proposals of new and supposedly extended
forms of Boltzmann entropy, apparently necessary to deal with \textit{%
nonextensive systems}. This does not seem to be correct \cite{LVR02b,Cha02},
and we are facing here the question posted before that the lack of
specificity has resulted in a plethora of expressions purporting to
represent the entropy of thermodynamics \cite{GC97}. The propagation of this
idea about a more general and universal expression for entropy, is a result
of the confusion arising from the fact that the proposed ones are \textit{%
informational entropies} (that is, generating functionals of probability
distributions): They\ provide a practical and useful tool for handling
problems where the researcher does not have access to the complete necessary
information on the relevant -- for the problem in hands -- characteristics
of the system, for properly applying the well established Boltzmann-Gibbs
quite general, and physically and logically sound, theory. That is, it
provides a way out to contour the failure we do have to satisfy Fisher's
\textit{Criterion of Sufficiency} in statistics of 1922 \cite{Fis22}, as
shown as we proceed. There are infinitely-many possible generating
functionals that can be introduced for that purpose, what was done around
the middle of past century by a school of statisticians. This lead to the
development of unconventional statistics, which are useful for making
predictions in the cases when the general and extremely successful
Boltzmann-Gibbs statistics has its application impaired once, as noticed,
the researcher cannot satisfy \textit{Fisher's Criterion of Sufficiency }
\cite{Fis22},\textit{\ that is, we are unable to introduce in the
calculations a proper description of the characteristics of the system that
are relevant for the experiment under consideration}.

As said a large number of possibilities (in principle infinitely-many) can
be explored, and Peter Landsberg quite properly titled an article \textit{%
Entropies Galore! }\cite{Lan99}. An infinite family is the one that can be
derived from Csiszer's general measure of cross-entropies (see for example
\cite{KK92}); other family has been proposed by Landsberg \cite{LV98}; and
specific informational entropies are, among others, the one of Skilling \cite
{Ski86} -- which can be used in mathematical economy --, and of Kaniadakis
\cite{Kan01} who used it in the context of special relativity \cite{Kan02}.
They, being generating functionals of probability distributions, give rise
to particular forms of statistics as the one of next section which we have
dubbed \textit{Unconventional Statistical Mechanics} \cite{LVR03a,VRL03a};
we do also have so-called \textit{Superstatistics} proposed by C. Beck and
E. G. D. Cohen for dealing with nonequilibrium systems with a stationary
state and intensive parameter fluctuations\ \cite{BC02,Bec03}; what can be
called \textit{Kappa Statistics} \cite{Kan01,Kan02}, and so on.

What we do actually have behind this question is the possibility to
introduce of such sophisticated method in physics, more precisely in
statistical mechanics, what we do try to describe in continuation.

\section{UNCONVENTIONAL STATISTICAL MECHANICS}

We begin calling the attention to the fact that Statistical Mechanics of
many-body systems has a long and successful history. The introduction of the
concept of probability in physics originated mainly from the fundamental
essay of Laplace of 1825 \cite{Lap25}, who incorporated and extended some
earlier seminal ideas (see for example \cite{Jay78}). As well known,
Statistical Mechanics attained the status of a well established discipline
at the hands of Maxwell, Boltzmann, Gibbs, and others, and went through some
steps related to changes, not in its fundamental structure, but just on the
substrate provided by microscopic mechanics. Beginning with classical
dynamics, statistical mechanics incorporated -- as they went appearing in
the realm of Physics -- relativistic dynamics and quantum dynamics. Its
application to the case of systems in equilibrium proceeded rapidly and with
exceptional success: equilibrium statistical mechanics gave -- starting from
the microscopic level -- foundations to Thermostatics, and the possibility
to build a Response Function Theory. Applications to nonequilibrium systems
began, mainly, with the case of local equilibrium in the linear regime
following the pioneering work of Lars Onsager \cite{Ons31} (see also \cite
{Cas45}).

For systems arbitrarily deviated from equilibrium and governed by nonlinear
kinetic laws, the derivation of an ensemble-like formalism proceeded at a
slower pace than in the case of equilibrium, and somewhat cautiously, with a
long list of distinguished scientists contributing to such development. It
can be noticed that Statistical Mechanics gained in the fifties an
alternative approach sustained on the basis of Information Theory \cite
{Jay78,Jay57a,Jay57b,Jay83,Jay89,Jay91,Jay93,Gra87,Gra88}: It invoked the
ideas of Information Theory accompanied with ideas of scientific inference
\cite{Jef61,Jef73}, and a variational principle (the latter being Jaynes'
principle of maximization of informational uncertainty -- also referred-to
as informational-entropy -- and called \textit{MaxEnt} for short),
compounding from such point of view a theory dubbed as \textit{Predictive
Statistical Mechanics} \cite
{Jay78,Jay57a,Jay57b,Jay83,Jay89,Jay91,Jay93,Jay86}. It should be noticed
that this is not a new paradigm in Statistical Physics, but a quite useful
and practical variational method which codifies the derivation of
probability distributions, which can be obtained by either heuristic
approaches or projection operator techniques \cite{LVR02a,LV90,LVR1,ZMR96}.
It is particularly advantageous to build nonequilibrium statistical
ensembles, as done here, when it systematizes the relevant work on the
subject that renowned scientists provided along the past century. The
informational-based approach is quite successful in equilibrium and near
equilibrium conditions \cite{Jay57a,Jay57b,Gra87,Gra88}, and in the last
decades has been, and is being, also applied to the construction of a
generalized ensemble theory for systems arbitrarily away from equilibrium
\cite{LVR02a,LV90,LVR1,ZMR96}. The nonequilibrium statistical ensemble
formalism (NESEF for short) provides mechanical-statistical foundations to
irreversible thermodynamics (in the form of Informational Statistical
Thermodynamics -- IST for short \cite{Hob66a,Hob66b,GCVL94,LVR2}), a
nonlinear quantum kinetic theory \cite{LV90,LVR1,LVL90} and a response
function theory \cite{LVR1,LV80} of a large scope for dealing with many-body
systems arbitrarily away from equilibrium. NESEF has been applied with
success to the study of a number of nonequilibrium situations in the physics
of semiconductors (see for example the review article of Ref. \cite{AVL92})
and polymers \cite{MVL98d}, as well as to studies of complex behavior of
boson systems in, for example, biopolymers (e.g. Ref. \cite{FMVL00}). It can
also be noticed that the NESEF-based nonlinear quantum kinetic theory
provides, as particular limiting cases, far-reaching generalizations of
Boltzmann \cite{RVL95}, Mori (together with statistical foundations for
Mesoscopic Irreversible Thermodynamics \cite{DCVJL96}) \cite{MVLCVJ98a}, and
Navier-Stokes \cite{RVL01} equations and a, say, Informational Higher-Order
Hydrodynamics, linear \cite{JCVVML02} and nonlinear \cite{RVL03}.

NESEF is built within the scope of the variational method on the basis of
the maximization of the informational-entropy in
Boltzmann-Gibbs-Shannon-Jaynes sense, that is, the average of minus the
logarithm of the time-dependent -- i.e. depending on the irreversible
evolution of the macroscopic state of the system -- nonequilibrium
statistical operator. It ought to be further emphasized that \textit{%
informational-entropy} -- a concept introduced by Shannon -- is in fact the
quantity of uncertainty of information, and \textit{has the role of a
generating functional for the derivation of probability distributions} (for
tackling problems in Communication Theory, Physics, Mathematical Economics,
and so on). There is \textit{one and only one} situation when Shannon-Jaynes
informational-entropy coincides with the true \textit{physical entropy of
Clausius in thermodynamics}, namely, the case of strict equilibrium; e.g.
\cite{Jay65,Lan99,JA02,Zub74}. For short, we shall refer to
informational-entropy as \textit{infoentropy.} As already noticed the
variational approach produces the well established equilibrium statistical
mechanics, and is providing a satisfactory formalism for describing
nonequilibrium systems in a most general form. This \textit{Boltzmann-Gibbs
Statistical Mechanics properly describes the macroscopic state of condensed
matter systems}, being a well established one and logically and physically
sound, but in some kind of situations, for example, involving
nanometric-scale systems with some type or other of fractal-like structures
or systems with long-range space correlations, or particular long-time
correlations, it becomes difficult to apply because of a \textit{deficiency
in the proper knowledge of the characterization of the states of the system}
in the problem one is considering (at either the microscopic or/and
macroscopic or mesoscopic level). This is, say, a practical difficulty (a
limitation of the researcher) in an otherwise general and highly successful
physical theory.

In fact, in a classical and fundamental paper of 1922 \cite{Fis22} by Sir
Ronald Fisher, titled ``On the Mathematical Foundations of \ Theoretical
Statistics'', are presented the basic criteria that a statistics should
satisfy in order to provide valuable results, that is, reliable predictions.
In what regards present day Statistical Mechanics in Physics two of them are
of major relevance, namely the \textit{Criterion of Efficiency} and the
\textit{Criterion of Sufficiency}. This is so because of particular
constraints that impose recent developments in physical situations involving
small systems (nanotechnology, nanobiophysics, quantum dots and
heterostructures in semiconductor devices, one-molecule transistors,
fractal\ electrodes in microbatteries, and so on), where on the one hand the
number of degrees of freedom entering in the statistics may be small, and on
the other hand boundary conditions of a fractal-like character are present
which strongly influence the properties of the system, what makes difficult
to introduce sufficient information for deriving a proper Boltzmann-Gibbs
probability distribution. Other cases when sufficiency is difficult to
satisfy is the case of large systems of fluids whose hydrodynamic motion is
beyond the domain of validity of the classical standard approach. It is then
required the use of a nonlinear higher-order hydrodynamics, eventually
including correlations and other variances (a typical example is the case of
turbulent motion). Also we can mention other cases where long-range
correlations have a relevant role (e.g. velocity distribution in clusters of
galaxies at a cosmological size, or at a microscopic size the already
mentioned case of one-molecule transistors where Coulomb interaction between
carriers is not screened and then of long range creating strong correlations
in space).

Hence, we may say that the proper use of Boltzmann-Gibbs statistics is
simply impaired because of either a great difficulty to handle the required
information relevant to the problem in hands, or incapacity on the part of
the researcher to have a correct access to such information. Consequently,
out of practical convenience or the force of circumstances, respectively, a
way to circumvent this inconveniency in such kind of ``anomalous''
situations, consists to resort to the introduction of \textit{modified forms
of the informational-entropy}, that is, other than the quite general one of
Shannon-Jaynes which leads to the well established statistics of
Boltzmann-Gibbs. These modified infoentropies which are built in terms of
the \textit{deficient characterization one does have of the system,} are
dependent on parameters -- called information-entropic indexes, or \textit{%
infoentropic indexes} for short with the understanding that refer to the
infoentropy. We reiterate the fundamental fact \textit{that these
infoentropies are generating functionals for the derivation of probabilities
distributions, and are not at all to be confused with the physical entropy
of the system}.

As already noticed, this alternative approach originated in the decades of
the 1950's and 1960's at the hands of theoretical statisticians, being
extensively used in different disciplines (economy, queueing theory,
regional and urban planning, nonlinear spectral analysis, and so on). Some
approaches were adapted for use in physics, and we present here an overall
picture leading to what can be called \textit{Unconventional Statistical
Mechanics} (USM for short and fully described in Refs. \cite{LVR03a,VRL03a}%
), consisting, as said before, in a way to patch the lack of knowledge of
characteristics of the physical system which are relevant for properly
determining one or other property (see also P. T. Landsberg in Refs. \cite
{Lan99} and \cite{LV98}) thus impairing the correct use of the conventional
one.

Use of the variational MaxEnt for building NESEF provides a powerful,
practical, and soundly-based procedure of a quite broad scope, which is
encompassed in what is sometimes referred-to as \textit{%
Informational-Entropy Optimization Principles} (see for example Ref. \cite
{KK92}). To be more precise we should say \textit{constrained optimization},
that is, restricted by the constraints consisting in the available
information. Such optimization is performed through calculus of variation
with Lagrange's method for finding the constrained extremum being the
preferred one.

Jaynes' variational method of maximization of the informational-statistical
entropy is connected -- via information theory in Shannon-Brillouin style --
to a principle of maximization of uncertainty of information. This is the
consequence of resorting to a principle of scientific objectivity \cite
{Jef73,Jef61}, which can be stated as: \textit{Out of all probability
distributions consistent with a given set of constraints, we must take the
one that has maximum uncertainty. }Its use leads to the construction of a
Gibbs' ensemble formalism, recovering the traditional one in equilibrium
\cite{Jay57a,Jay57b,Gra87}, and allowing for the extension to systems far
from equilibrium \cite{Gra88,LVR02a,LV90,LVR1,ZMR96}.

Jaynes' MaxEnt is a major informational-entropy optimization principle
requiring, as noticed, that we should use only the information which is
accessible but scrupulously avoiding to use information not proven to be
available. This is achieved by maximizing the uncertainty that remains after
all the given information has been taken care of.

Jaynes' MaxEnt aims at maximizing uncertainty when subjected to a set of
constraints which depend on each particular situation (given values of
observables and theoretical knowledge). But uncertainty can be a too deep
and complex concept for admitting a unique measure under all conditions: We
may face situations where uncertainty can be associated to different \textit{%
degrees of fuzziness in data and information}. As already noticed, this is a
consequence, in Statistical Mechanics, of a lack of a proper description of
the physical situation. This corresponds to being violated the \textit{%
Criterion of Sufficiency in the characterization of the system} (``the
statistics chosen should summarize the whole of the \textit{relevant}
information supplied by the sample'') \cite{Fis22}\textit{.} This could
occur at the level of the microscopic dynamics (e.g. lack of knowledge of
the proper eigenstates, all important in the calculations), or at the level
of macroscopic dynamics (e.g. when we are forced, because of deficiency of
knowledge, to introduce a low-order truncation in the higher-order
hydrodynamics that the situation may require). Hence, in these\
circumstances it may arise the necessity of introducing alternative types of
statistics, with the accompanying \textit{indexed (or structural)
informational-entropies}, (\textit{infoentropies} for short\textit{)}
different of the quite general and well established one of Boltzmann-Gibbs,
as it has been discussed in the first section\textit{.} \bigskip \newpage

{\LARGE TABLE I: Informational-Statistical Entropies}

\bigskip

{\Large Conventional (Universal) ISE}

\begin{tabular}{ll}
Boltzmann-Gibbs-Shannon-Jaynes ISE & $\ \left\{
\begin{array}{c}
-Tr\left\{ \varrho \ln \varrho \right\}
\end{array}
\right. $ \\
(from K\"{u}lback-Leibler measure) &
\end{tabular}

\bigskip

{\Large Unconventional (entropic-index-dependent) ISEs}

\begin{tabular}{ll}
From Havrda-Charvat measure\ \ \ \ \ \ \  & $\left\{
\begin{array}{c}
-\frac{1}{\alpha -1}Tr\left\{ \varrho ^{\alpha }-\varrho \right\} \\
\ \alpha >0\ and\ \alpha \neq 1
\end{array}
\right. $ \\
&  \\
From Sharma-Mittal measure & $\left\{
\begin{array}{c}
-\frac{W^{\beta -1}}{\alpha -\beta }Tr\left\{ \left[ W^{\alpha -\beta
}\varrho ^{\alpha -\beta +1}-\varrho \right] \varrho ^{\beta -1}\right\} \\
\alpha >1,\ \beta \leq 1\ or\ \alpha <1,\ \beta \geq 1
\end{array}
\right. $ \\
&  \\
From Renyi measure & $\left\{
\begin{array}{c}
-\frac{1}{\alpha -1}\ln Tr\left\{ \varrho ^{\alpha }\right\} \\
\ \alpha >0\ and\ \alpha \neq 1
\end{array}
\right. $ \\
&  \\
From Kapur measure & $\left\{
\begin{array}{c}
-\frac{1}{\alpha -\beta }\left[ \ln Tr\left\{ \varrho ^{\alpha }\right\}
-\ln Tr\left\{ \varrho ^{\beta }\right\} \right] \\
\alpha >0,\ \beta >0\ and\ \alpha \neq \beta
\end{array}
\right. $%
\end{tabular}

\bigskip

Applying MaxEnt to any of these unconventional infoentropies we obtain a
probability distribution deemed appropriate for the given problem in hands,
namely, either the conventional probability distribution when Shannon-Jaynes
infoentropy is used, or the so-called in Pearsons' nomenclature \textit{%
heterotypical probability distributions} when other infoentropies are used.
There are infinitely-many of these infoentropies, which, as said, are
dependent on one or more parameters\ (the infoentropic indexes), and in
Table \textbf{I} we list four of these informational-statistical entropies
(ISE), plus the Shannon-Jaynes one (of Boltzmann-Gibbs type) which is
parameter free as it should for being the most general one \cite
{KK92,LVR03a,LVR1}. In Table \textbf{I} $W^{-1}$ is the constant value of
probability in the uniform distribution, and $\alpha $ and $\beta $
infoentropic indexes (open parameters).

Renyi's approach appears to be a particularly convenient one for dealing
with fractal systems as discussed in Ref. \cite{JA02}, where it is pointed
out that predictions obtained resorting to the approach of maximization in
Shannon-Jaynes approach including fractality can be equivalently obtained
using Renyi's approach ignoring fractality. Renyi's ISE has been studied by
Takens and Verbitski \cite{TV98}, and a variation of it is
Hentschel-Procaccia infoentropy \cite{HP83}. For the Havrda-Charvat
structural $\alpha $-entropy, one akin to the case $\alpha =2$ has been
considered by I. Prigogine in connection with practical and theoretical
difficulties with Boltzmann ideas when extending them from the dilute gas to
dense gases and liquids \cite{Pri80}. Prigogine argues that to cope with
such situations one would need a statistical expression of entropy that
depends explicitly on \textit{correlations}, as is the case of the
Havrda-Charvat structural $\alpha $-entropy for $\alpha =2$ (also in the
case of Renyi infoentropy).

It can be noticed that taking $\beta =1$ reduces Kapur ISE to the one of
Renyi, and Sharma-Mittal ISE to the one of Havrda-Charvat. Moreover, taking
also $\alpha =1$, is obtained an ISE which is of the \textit{form} of
Boltzmann-Gibbs-Shannon-Jaynes ISE. What we do have in these ISE's, or in
any other one of the infinitely-many which are possible, is that when the
adjustment of the parameters (the infoentropic indexes) on which they depend
-- let it be in a calculation or as a result of the comparison with the
experimental data -- produces Boltzmann-Gibbs result, this gives an
indication that the principle of sufficiency is being satisfied, i.e., for
such particular situation the description of the system we are doing
includes all the \textit{relevant} characterization that properly determines
the physical property that is measured in the \textit{given} experiment
being analyzed.

Moreover, we again stress the fundamental fact that the structural
informatio-\linebreak nal-entropies (quantity of uncertainty of information)
are not to be confused with the Clausius-Boltzmann physical entropy: There
is one and only one case when there is an equivalence, consisting of Shannon
infoentropy when the system is strictly in equilibrium \cite
{Jay65,Lan99,JA02}. Boltzmann-Gibbs-Shannon-Jaynes nonequilibrium
informational entropy and its role in NESEF is extensively discussed in
Refs. \cite{LVR1,LVR2,LVR00}.

It is quite relevant to notice that for each kind of statistical entropy it
is necessary in an \textit{ad hoc} manner, to introduce definitions of
average values of observables with particular forms, what is required to
obtain \textit{a posteriori} consistent results. For the case of
Kullback-Leibler measure, or Shannon-Jaynes statistical
informational-entropy, we must use the usual expression, i.e. the average of
quantity $\hat{A}$ is given by
\begin{equation}
\left\langle \hat{A}\right\rangle =Tr\left\{ \hat{A}\varrho \right\} \qquad ,
\label{eqA4a}
\end{equation}
while for the case of Renyi ISE, needs be introduced an average of the form
\begin{equation}
\left\langle \hat{A}\right\rangle =Tr\left\{ \hat{A}\mathcal{D}_{\gamma
}\left\{ \varrho \right\} \right\} \qquad ,  \label{eqA4b}
\end{equation}
that is, in terms of the so-called \textit{escort probability} of order $%
\gamma $\ \cite{Ren70,BS93}
\begin{equation}
\mathcal{D}_{\gamma }\left\{ \varrho \right\} =\varrho ^{\gamma }/Tr\left\{
\varrho ^{\gamma }\right\} \qquad ,  \label{eqA4c}
\end{equation}
which is also the one to be used in the case of Havrda-Charvat statistics. A
detailed discussion on the role of escort probability is given in Ref. \cite
{LVR03a}.

We call the attention to the fact that \textit{USM is to be based on the use
of both definitions, namely, the heterotypical probability distribution and
the escort probability} (notice that for probability distributions other
than Renyi and Havrda-Charvat other definitions of escort probabilities
should be introduced). The role of the escort probability accompanying the
heterotypical-probability distribution is that both complement each other in
order to redefine, in the sense of weighting, the values of the
probabilities associated to the physical states of the system \cite{LVR03a}.

Let us now consider the use of Renyi informational entropy, i.e.
\begin{equation}
S_{\alpha }\left( t\right) =-\frac{1}{\alpha -1}\ln Tr\left\{ \ \left[ \bar{%
\varrho}_{\alpha }\left( t,0\right) \right] ^{\alpha }\right\} \qquad ,
\label{A12}
\end{equation}
\ in the case of a fluid of single molecules driven out of equilibrium and
in contact with ideal reservoirs; a more general situation is discussed in
\cite{LVR03a}.\ We recall that a recent application of Renyi's statistics
for dealing with (multi)fractal systems is presented by Jizba and Arimitzu
\cite{JA02}: there it is addressed the question on how Renyi's approach
appears as a quite convenient one in such cases. Further considerations on
Renyi's approach can be consulted in the articles by Hentschel and Procaccia
\cite{HP83} and Takens and Verbitski \cite{TV98}.

We proceed to derive the nonequilibrium statistical operator in Renyi's
approach, beginning to look for the ``instantaneously frozen'' auxiliary
statistical operator which, in the situation above described, consists of
the product of the one of the system, say $\bar{\varrho}_{\alpha }\left(
t,0\right) $, times the constant one of the thermal bath and ideal
reservoirs. Let us now look for $\bar{\varrho}_{\alpha }\left( t,0\right) $
which follows by maximizing $S_{\alpha }$ of Eq. (\ref{A12}) subjected to
the conditions of normalization
\begin{equation}
Tr\left\{ \bar{\varrho}_{\alpha }\left( t,0\right) \right\} =1\qquad ,
\label{eqA20}
\end{equation}
and the constraints consisting of the average values, say $Q_{j}\left(
\mathbf{r},t\right) $, as defined by Eq. (\ref{eqA4b}), of the basic
dynamical variables, say $\left\{ \hat{P}_{j}\left( \mathbf{r}\right)
\right\} $, namely
\begin{equation}
Q_{j}\left( \mathbf{r},t\right) =Tr\left\{ \hat{P}_{j}\left( \mathbf{r}%
\right) \stackrel{-}{\mathcal{D}}_{\alpha }\left\{ \bar{\varrho}\left(
t,0\right) \right\} \right\} \qquad ,  \label{eqA21}
\end{equation}
where
\begin{equation}
\stackrel{-}{\mathcal{D}}_{\alpha }\left\{ \bar{\varrho}\left( t,0\right)
\right\} =\left[ \bar{\varrho}_{\alpha }\left( t,0\right) \right] ^{\alpha
}\ /\ Tr\left\{ \left[ \bar{\varrho}_{\alpha }\left( t,0\right) \right]
^{\alpha }\right\}  \label{eqA22}
\end{equation}
is the corresponding escort probability \cite{Ren70,BS93} (cf. discussion
after Eq. (\ref{eqA4c}) above); it can be observed that the order of the
escort probability is the same infoentropic index in Renyi's informational
entropy \cite{Ren70}.

It follows that
\begin{equation}
\bar{\varrho}_{\alpha }\left( t,0\right) =\frac{1}{\bar{\eta}_{\alpha
}\left( t\right) }\left[ 1+\left( \alpha -1\right) \sum\limits_{j}\int
d^{3}r\ F_{j\alpha }\left( \mathbf{r},t\right) \ \Delta \hat{P}_{j}\left(
\mathbf{r},t\right) \right] ^{-\frac{1}{\alpha -1}},  \label{eqA23}
\end{equation}
where
\begin{equation}
\Delta \hat{P}_{j}\left( \mathbf{r},t\right) =\hat{P}_{j}\left( \mathbf{r}%
\right) -Q_{j}\left( \mathbf{r},t\right) \qquad ,  \label{eqA24}
\end{equation}
with $Q_{j}\left( \mathbf{r},t\right) $ given in Eq. (\ref{eqA21}),
\begin{equation}
\bar{\eta}_{\alpha }\left( t\right) =Tr\left\{ \left[ 1+\left( \alpha
-1\right) \sum\limits_{j}\int d^{3}r\ F_{j\alpha }\left( \mathbf{r},t\right)
\ \Delta \hat{P}_{j}\left( \mathbf{r},t\right) \right] ^{-\frac{1}{\alpha -1}%
}\right\} \qquad ,  \label{eqA25}
\end{equation}
ensures the normalization condition, and $F_{j\alpha }$ are\ the Lagrange
multipliers that the variational method introduces, which are related to the
basic variables through Eq. (\ref{eqA21}).

In terms of the auxiliary $\bar{\varrho}_{\alpha }$,\ the statistical
distribution is given by \cite{LVR02a,LV90,LVR1,ZMR96}
\begin{equation}
\varrho _{\alpha \epsilon }\left( t\right) =\epsilon \int\limits_{-\infty
}^{t}dt^{\prime }e^{\epsilon \left( t^{\prime }-t\right) }\bar{\varrho}%
_{\alpha }\left( t^{\prime },t^{\prime }-t\right) \qquad ,  \label{eqA26}
\end{equation}
where, we recall,
\begin{equation}
\bar{\varrho}_{\alpha }\left( t^{\prime },t^{\prime }-t\right) =\exp \left\{
-\frac{1}{i\hslash }\left( t^{\prime }-t\right) \hat{H}\right\} \bar{\varrho}%
_{\alpha }\left( t^{\prime },0\right) \exp \left\{ \frac{1}{i\hslash }\left(
t^{\prime }-t\right) \hat{H}\right\} \qquad ,  \label{eqA27}
\end{equation}
and $\epsilon $ is a positive infinitesimal that goes to zero after the
trace operation in the calculation of averages has been performed. The
statistical distribution of Eq. (\ref{eqA26}) satisfies the Liouville
equation
\begin{equation}
\frac{\partial }{\partial t}\varrho _{\alpha \epsilon }\left( t\right) +%
\frac{1}{i\hslash }\left[ \varrho _{\alpha \epsilon }\left( t\right) ,\hat{H}%
\right] =-\epsilon \left[ \varrho _{\alpha \epsilon }\left( t\right) -\bar{%
\varrho}_{\alpha }\left( t,0\right) \right] \qquad ,  \label{eqA28}
\end{equation}
with the presence of the infinitesimal source introducing Bogoliubov's
symmetry breaking procedure (quasiaverages) \cite{Bog70}, in the present
case the one of time reversal and in that way are discarded the advanced
solutions of the full Liouville equation. Thus, the retarded solutions have
been selected, and, \textit{a posteriori}, this is transmitted to the
kinetic equations producing a \textit{fading memory} and irreversible
behavior (cf. Refs. \cite{LVR1,LVL90}).

We also call the attention to the fact that for average values, as given by
Eq. (\ref{eqA4b}), we then have
\begin{equation}
\left\langle \hat{A}\right\rangle =Tr\left\{ \hat{A}\mathcal{D}_{\alpha
\epsilon }\left\{ \varrho _{\alpha \epsilon }\left( t\right) \right\}
\right\} \qquad ,  \label{A22}
\end{equation}
where
\begin{equation}
\mathcal{D}_{\alpha \epsilon }\left\{ \varrho _{\alpha \epsilon }\left(
t\right) \right\} =\varrho _{\alpha \epsilon }^{\alpha }\left( t\right)
/Tr\left\{ \varrho _{\alpha \epsilon }^{\alpha }\left( t\right) \right\}
\qquad ,  \label{A23}
\end{equation}
and it is implicit the limit $\epsilon \rightarrow 0$ after the calculation
of traces has been performed.

Because of the boundary condition $\varrho _{\alpha \epsilon }\left(
t_{o}\right) =\bar{\varrho}_{\alpha }\left( t_{o},0\right) $ ($%
t_{o}\rightarrow -\infty $) \cite{LVR02a,LV90,LVR1,ZMR96}, we have that $\
\mathcal{D}_{\alpha \epsilon }\left\{ \varrho _{\alpha \epsilon }\left(
t_{o}\right) \right\} =\stackrel{-}{\mathcal{D}}_{\alpha }\left\{ \bar{%
\varrho}_{\alpha }\left( t_{o},0\right) \right\} $, where $\stackrel{-}{%
\mathcal{D}}_{\alpha }$ is given by Eq. (\ref{eqA22}). For $\epsilon
\rightarrow 0$, $\varrho _{\alpha \varepsilon }$ satisfies a true Liouville
equation [cf. Eq. (\ref{eqA28})], and so does $\mathcal{D}_{\alpha \epsilon
} $, and we recall that the infinitesimal source on the right-hand side of
Eq. (\ref{eqA28}) is selecting the retarded solutions of the true Liouville
equation (via, then, Bogoliubov's method of quasiaverages, as previously
noticed). Hence, for the given initial condition and the imposition of
discarding the advanced solutions, $\mathcal{D}_{\alpha \epsilon }\left\{
\varrho _{\alpha \epsilon }\left( t\right) \right\} $ also satisfies a
modified Liouville equation, and we can write
\begin{equation}
\mathcal{D}_{\alpha \epsilon }\left\{ \varrho _{\alpha \epsilon }\left(
t\right) \right\} =\stackrel{-}{\mathcal{D}}_{\alpha }\left\{ \bar{\varrho}%
_{\alpha }\left( t,0\right) \right\} +\mathcal{D}_{\alpha \epsilon }^{\prime
}\left( t\right) \qquad ,  \label{A25}
\end{equation}
where $\stackrel{-}{\mathcal{D}}_{\alpha }\left\{ \bar{\varrho}_{\alpha
}\left( t,0\right) \right\} $ is given by Eq. (\ref{eqA22}), and
\begin{equation}
\mathcal{D}_{\alpha \epsilon }^{\prime }\left( t\right)
=-\int\limits_{-\infty }^{t}dt^{\prime }e^{\epsilon \left( t^{\prime
}-t\right) }\frac{d}{dt^{\prime }}\stackrel{-}{\mathcal{D}}_{\alpha }\left\{
\bar{\varrho}_{\alpha }\left( t^{\prime },t^{\prime }-t\right) \right\}
\qquad .  \label{A26}
\end{equation}

Introducing Eq. (\ref{A25}) into Eq. (\ref{A22}), we can see that the
averages are composed of an ``instantaneously frozen'' (at time $t$)
contribution, plus a contribution associated to the irreversible processes
and including historicity. For the basic dynamical quantities, and \textit{%
only} for them, it follows that
\begin{equation}
Q_{j}\left( \mathbf{r},t\right) =Tr\left\{ \hat{P}_{j}\mathcal{D}_{\alpha
\epsilon }\left\{ \varrho _{\alpha \epsilon }\left( t\right) \right\}
\right\} =Tr\left\{ \hat{P}_{j}\stackrel{-}{\mathcal{D}}_{\alpha }\left\{
\bar{\varrho}_{\alpha }\left( t,0\right) \right\} \right\} \qquad .
\label{A27}
\end{equation}
with, as already noticed, being implicit the limit of $\epsilon $ going to $%
+0$ to be taken after the calculation of the trace operation has been
performed \cite{LVR02a,LV90,LVR1,ZMR96}.

After the nonequilibrium distribution using a heterotypical index-dependent
information-\linebreak al-entropy has been derived, next step -- like done
in the conventional case \cite{LVR1,LVR2,LVL90,LV80,LVR00,LVR01} -- should
consists in deriving for arbitrarily far-from-equilibrium systems, a
nonlinear quantum kinetic theory, a response function theory, and, of
course, a systematic study of experimental results, that is, a full
collection of measurements of diverse properties of the system, amenable to
be studied in terms of structural (infoentropic-index dependent)
informational-entropies, what is fundamental for the validation of the
theory (see for example Refs. \cite{VRL03a,VGKRL02,VLML03,LVML03,VRLJCV03}).

\section{IRREVERSIBLE THERMODYNAMICS AND\protect\linebreak ENTROPY}

As know, nonequilibrium thermodynamics is involved with finite and
irreversible processes, and, for that reason, it is also denominated
Irreversible Thermodynamics. As a general rule special interest is currently
focused on the case of open systems, which are coupled to external sources
which supply them energy and matter. This manifests itself in macroscopic
changes of the thermodynamic variables which naturally become dependent on
space position and on time.

The principal characteristic of the processes involved, whether they are
stationary or time dependent, is that they evolve with a positive production
of internal entropy. Here we make contact with the second fundamental
principle of thermodynamics. According to Planck \cite{Pla}, ``The second
law of thermodynamics is essentially different from the first law, since it
deals with a question in no way touched upon by the first law, viz., the
direction in which a process takes place in nature''. The second law has
several equivalent formulations: the one due to Clausius refers to heat
conduction, and, again according to Planck \cite{Pla}, it can be expressed
as ``heat cannot by itself pass from a cold to a hot body. As Clausius
repeatedly and expressly pointed out, this principle does not merely say
that heat does not flow from a cold to a hot body -- that is self-evident,
and is a condition of the definition of temperature -- but it expressly
states that heat can in no way and by no process be transported from a
colder to a warmer body without leaving further changes, i.e. without
compensation''.

The concept of irreversibility clearly appears in Clausius' work, who can be
considered the founder of Thermodynamics as an autonomous and unified
science. His presentation of the second law appears precisely as a criterion
for evolution, governing irreversible behavior. It implies the definition
and use of the extremely difficult concept of the title, namely, \textit{%
entropy}. Using the usual notation, for an infinitesimal process the state
function entropy $S$ suffers a modification $dS=\frak{d}Q/T$ if the process
is reversible, while for irreversible processes there follows $dS>\frak{d}%
Q/T $, where $\frak{d}Q$ is the heat exchange in the process and $T$ the
absolute temperature, and $\frak{d}$ stands for a nonexact differential
(thus, $1/T$ acts as an integrating factor). The last inequality can be
expressed in the form of the balance equation
\begin{equation}
dS=\frac{\frak{d}Q}{T}+\frac{\frak{d}Q^{\prime }}{T}  \label{eq1}
\end{equation}
with $\frak{d}Q^{\prime }>0.$ In this form of the second law, the quantity $%
\frak{d}Q^{\prime }$, called by Clausius \textit{uncompensated heat}, can be
considered as a first evaluation of the degree of irreversibility of a
natural process.

Equilibrium thermodynamics describes states of matter that are greatly
privileged: Planck \cite{Pla} has emphasized that the second law
distinguishes among the several types of states in nature, some of which act
as attractors to others: Irreversibility is an expression of this
attraction. For systems together with the reservoirs to which they are
connected there is an attraction towards equilibrium (thermodynamic
potentials at a minimum value compatible with the constraints imposed by the
reservoirs). In nonequilibrium systems, while in a stationary state in a
linear regime near equilibrium, the attractor is the state of minimum
internal entropy production \cite{LVR2}.

According to Eq. (\ref{eq1}), the change of entropy is composed of two
separate types of contributions: the term $\frak{d}Q/T$ related to the
exchange of heat with the surroundings, expressed as $d_{e}S$, and the term $%
\frak{d}Q^{\prime }/T$, due exclusively to the irreversible processes that
develop in the interior of the system, expressed as $d_{i}S$ so we can write
\begin{equation}
dS=d_{e}S+d_{i}S,\qquad with\text{ \qquad }d_{i}S\geq 0.  \label{eq2}
\end{equation}
Equation (\ref{eq2}) provides the framework for an entropy equation of
balance, equivalent to Eq. (\ref{eq1}). Let us emphasize that the term $%
d_{e}S$ involves all the contributions resulting from exchanges (energy,
matter, etc.) with the environment.

After what has been manifested in Section \textbf{1}, it is certainly a
truism to say that entropy has a very special status in physics, expressing,
in a very general way, the tendency of physical systems to evolve in an
irreversible way, characterizing the eventual attainment of equilibrium, and
given a kind of measurement of the order that prevails in the system. It is
important to further stress that it is a very well established concept in
equilibrium situations, however requiring an extension and clear
comprehension in the case of open systems, mainly in far-from-equilibrium
conditions, a question that remains controversial. On the basis of the use
of entropy as a state function, the properties of systems in \textit{%
equilibrium} are very well described when in conjunction with the two
fundamental laws. Let us add some additional consideration to what have been
said in Section \textbf{1.}

Lawrence Sklar \cite{Skl93} has noted that the concept of entropy is the
most purely thermodynamic concept of all, and J. Bricmont \cite{Bri96} has
commented that there is some kind of mystique about entropy. Again according
to Sklar, given the abstractness of entropy and its high place up in the
theory as well as its unrelatedness to immediate sensory qualities or
primitive measurements (as temperature is related to these), it is not
surprising than in seeking the statistical mechanical correlate of \textit{%
nonequilibrium thermodynamic entropy} we have the least guidance from the
surrounding embedding theory. Exists an openness in what to choose as the
surrogate for entropy. Thus, it should be expected to arise a wide variety
(plethora) of ``entropies'', each functioning well for the specific purposes
for which it was introduced. J. Meixner \cite{Mei73} asks: \textit{Is the
concept of nonequilibrium entropy superfluous?}, for in continuation to
comment that one is so much accustomed to the concept of entropy that one
would like to retain it as a quantity of physical significance. He also
points to the difficulty of a definition if one does not have a clearly
defined physical state of the system. This is the main difficulty, as also
pointed out by Bricmont in\ that \textit{we may define as many entropies as
we can find sets of macroscopic variables}. Also, with the coarse-graining
procedure there is not a sharp distinction between microscopy and
macroscopy, passing through mesoscopy, so that we can arrive at many values
of the say ``entropy'', including arriving at the zero value when a complete
microscopic description is given, and we have a pure mechanical description
and no thermodynamics exists (\cite{Mei73,Mei74}, see also Jaynes in \cite
{Jay65}). Jaynes rightly says that he does not know what is the entropy of a
cat; the problem being that we are unable to precisely define a set of
macrovariables that properly specify the thermodynamic state of the cat.

At this point it is worth to present, with some modifications, several quite
appropriate remarks made by Bricmont in the cited reference, which we
roughly summarize here:

(i) These entropies are not subjective but objective as are the
corresponding macroscopic variables. Called ``anthropomorphic'' by Jaynes
following Wigner, they may be referred to as ``contextual'', i.e. they
depend on the physical situation and on its level of description.

(ii) The ``usual'' or ``traditional''\ \textit{entropy of Clausius}
corresponds to the particular choice of macroscopic variables for a free
monatomic gas in \textit{equilibrium} (energy, specific volume, number of
particles). The derivative with respect to the energy of \textit{that}
entropy defines the reciprocal of the Kelvin's absolute\ temperature.

(iii) The second law in the form ``Entropy increases'' becomes undetermined:
which entropy? Between two states of equilibrium \ is the Clausius' one.
Otherwise is not clear.

(iv) Whichever the chosen functional form for an entropy, in most cases is
hard to compute or estimate. One needs to begin with the equations of
evolution for the chosen basic variables and solve them for appropriate
initial and boundary conditions. Moreover, irreversibility as characterized
by some $\mathcal{H}$-theorem -- as the one of Boltzmann -- does not
directly relate the $\mathcal{H}$-function to whatever may be the entropy.
It only ensures that the choice of the initial condition and some \textit{ad
hoc} nonmechanical hypotheses (a Stosszahlanzatz) gives a time-arrow and
relaxation towards final equilibrium.

(v) There is no difficulty with Liouville theorem of invariance of extension
in phase\ space; this is a purely mechanical result. At the statistical
level the extent of the volume of space points, compatible with the
macroscopic (or mesoscopic) description, changes because these constraints
change in time, and the characteristic set of microscopic points changes.
The evolution of such set is a different thing that the set of trajectories
of given points in a volume of phase space, whose volume is indeed conserved
according to Liouville theorem. Moreover,

(vi) It ought to be noticed that Gibbs' entropy is in fact constant in time,
because NESEF conserves the initial information, being then the said
fine-grained entropy. But, as Fig. \textbf{1} shows, as time elapses it \
gets outside the informational subspace and therefore is no longer
describing the state of the system in terms of the chosen set of basic
variables. It needs be projected at each time on the informational subspace,
again as shown in Fig. \textbf{1}, thus introducing loss of information. Of
course equilibrium is a particular case when Gibbs' entropy coincides with
Clausius thermodynamic entropy, and in the NESEF formalism it remains always
at the point of initial preparation (which is the one\ of equilibrium).

(vii) Bricmont makes a similar statement to that of Meixner: \textit{Why
should one worry so much about entropy for nonequilibrium states}? To
account for the irreversible behavior of the macroscopic variables is not
necessary to introduce some kind of entropy function that evolves
monotonically in time. It is not required to account for irreversibility,
however it may be interesting or useful to do so. In the case we have
presented of informational entropies applies this question of interest and
usefulness: it allows for a better clarification of the meaning and
interpretation of the Lagrange multipliers; to introduce the production of
informational entropy and the derivation of useful criteria for evolution
and stability; \ to better characterize the dissipative processes that
develops in the media (organizing them in increasing orders of covariances
of the informational-entropy production operator); to better analyze
fluctuations out of equilibrium and work out studies on complementarity of
micro/macro descriptions; and so on (see Ref. \cite{LVR2} for details).

Therefore, in principle and to all appearances, a true thermodynamic entropy
is only clearly defined, via Clausius approach, in strictly equilibrium
conditions. Out of equilibrium, quasi-entropies (in our nomenclature) may be
introduced and be of utility, but it needs be clearly stated which is the
definition, and to devise a particular name characterizing each: say,
entropy in Classical (Onsagerian) Irreversible Thermodynamics, entropy in
Extended Irreversible Thermodynamics, NESEF-entropy of Informational
Statistical Thermodynamics, and so on and so forth \cite{LVR2}.

The informational-entropy, to be related to the one in phenomenological
theories, depends on a chosen set of basic variables and arises from a
\textit{projected part} of the logarithm of the said nonequilibrium
statistical operator $\ln \varrho _{\varepsilon }\left( t\right) $, which is
$\ln \overline{\varrho }\left( t,0\right) $ (as shown in Fig. \textbf{1}).
This process is a \textit{coarse-grained-type procedure which restricts us
to have as only accessible microstates in phase space those in the subspace
spanned by the chosen basic dynamic variables}. Irreversible effects are
contained in the complementary part of the nonequilibrium statistical
operator, namely $\varrho _{\varepsilon }^{\prime }$, as it is demonstrated
by the proven generalized $\mathcal{H}$-theorem presented in Ref. \cite{LVR2}%
.

Furthermore, the local informational-entropy production function is
predominantly positive definite, but in any case these results cannot be
connected with the formulation of the second law, until a clear cut
definition of the entropy function in nonlinear thermodynamics for systems
arbitrarily away from equilibrium is obtained. This statement has to be
further clarified since here the second law must necessarily be understood
as some extension of Clausius formulation, with the latter, we recall,
involving changes between two equilibrium states and $S$ is the calorimetric
entropy which is uniquely defined. We emphasize once more that in
nonequilibrium states it is very likely that many different definitions of
surrogates of the entropy are feasible depending essentially on how to
obtain -- in some sense -- a complete set of macrovariables \ that may
unequivocally characterize the macrostate of the system under the given
experimental conditions, and thus agreeing with Meixner conjecture \cite
{Mei73}. \ In the Informational Statistical Thermodynamics -- which can be
considered as given microscopic, i.e. mechano-statistical foundations, to
macroscopic Extended Thermodynamics --, the associated informational entropy
(or quasi-entropy) has been discussed in detail in Refs. \cite{LVR2,LVR1},
where are presented the derivation: of nonequilibrium equations of state; of
a generalized $\mathcal{H}$-theorem; of evolution and (in)stability
criteria; of a generalized Clausius relation; of fluctuations and
Maxwell-like relations; of a Boltzmann-like relation: $\bar{S}\left(
t\right) =k_{B}\ \ln W\left( t\right) $; and comparison of theory and
experimental results is made.

As a results of this, it may be stated that NESEF seems to offer a sound
formalism to give foundations to irreversible thermodynamics on a
statistical-mechanical basis, an approach providing what has been
referred-to as Informational Statistical Thermodynamics. Thus, for the case
of systems under quite arbitrary dissipative conditions (no restriction to
local equilibrium, linearity, etc.) a theoretical treatment of a very large
scope follows for the thermodynamics, transport properties, and response
functions of nonequilibrium systems. Paraphrasing Zwanzig \cite{Zwa81}, we
remark that, seemingly, NESEF possesses a remarkable compactness and has by
far a most appealing structure, being a very effective method for dealing
with nonlinear and nonlocal in space and time transport processes in
far-from-equilibrium many-body systems.

\section{COMMENTS AND CONCLUDING REMARKS}

Summarizing, we first notice the relevant point that in the construction of
a statistical mechanics, the derivation of an appropriate (for the problem
in hands) probability distribution -- associated to a set of constraints
imposed on the system -- can be obtained in a compact and practical way by
means of optimization (variational) principles in a context related to
information theory. These are methods of maximization of the so-called
informational-entropies (better called quantities of uncertainty of
information) or minimization of distances in a space of probability
distributions (MaxEnt and MinxEnt respectively) \cite{KK92}.

In the original formulation of Shannon and Jaynes use was made of
Boltzmann-Gibbs statistical-entropy, which in MaxEnt provides the
canonical-like (exponential) distributions of classical, relativistic, and
quantum statistical mechanics. In Ref. \cite{LVR1} it is described its use
for the case of many-body systems arbitrarily far removed from equilibrium,
and the discussion of the dissipative phenomena that unfold in such
conditions (mainly ultrafast relaxation processes; see Ref. \cite{AVL92}).

This approach has been exceedingly successful in conditions of equilibrium,
and is a very promising one for nonequilibrium conditions. To have a
reliable statistical theory in these situations is highly desirable since in
very many situations -- as for example are the case of electronic and
optoelectronic devices, chemical reactors, fluid motion, and so on -- the
system is working in far-from-equilibrium conditions.

However the enormous success and large application of Shannon-Jaynes method
to\linebreak Laplace-Maxwell-Boltzmann-Gibbs statistical foundations of
physics, as it has been noticed, some cases look as difficult to be properly
handled within the Boltzmann-Gibbs formulation, as a result of existing some
kind of fuzziness in data or information, that is, the presence of a
condition of insufficiency in the characterization of the (microscopic
and/either macroscopic or mesoscopic) state of the system. Such, say,
difficulty with the proper characterization of the system in the problem in
hands, (which is a practical one and, we stress, not intrinsic to the
\textit{most general and complete Boltzmann-Gibbs formalism}) can be, as
shown, patched with the introduction of peculiar parameter-dependent
alternative structural informational-entropies (see Table \textbf{I}).

Particularly, to deal with systems with some kind of fractal-like structure
the use of Boltzmann-Gibbs-Shannon-Jaynes infoentropy would require to
introduce as information the highly correlated conditions that are in that
case present. Two examples in condensed matter physics are ``anomalous''
diffusion \cite{VGKRL02} and ``anomalous'' optical spectroscopy \cite{VLML03}%
, when fractality enters via the non-smooth topography of the boundary
surfaces which have large influence on phenomena occurring in constrained
geometries (nanometer scales in the active region of the sample). In the
conventional approach, the spatial correlations that the granular boundary
conditions introduce need be given as information (to satisfy the criterion
of sufficiency, since they are quite relevant for determining the behavior
of the system in the nanometric scales involved), but to handle them is
generally a nonfeasible task. For example, in the second case above
mentioned one has no easy access to the determination of the detailed
topography of the surfaces which limit the active region of the sample (the
nanometric quantum wells in semiconductor heterostructures), what can be
done in the first case using atomic-force microscopy and the determination
of the fractal dimension involved is possible. Hence the most general and
complete Boltzmann-Gibbs formalism in Shannon-Jaynes approach becomes
hampered out and is difficult to handle, and then, as shown, use of other
types of informational-entropies (better called generating functionals for
deriving probability distributions) may help to circumvent such
inconveniency by introducing alternative algorithms (dependent on the
so-called informational-entropic indexes), that is, the derivation of
heterotypical probability distributions on the basis of the constrained
maximization of unconventional informational-statistical entropies (quantity
of uncertainty of information), to be accompanied, as noticed in the main
text, with the use of the so-called escort probabilities.

In brief, we recall that, \textit{Unconventional Statistical Mechanics
consists of two steps: }$1$\textit{.} \textit{The choice of a deemed
appropriate structural informational-entropy for generating the
heterotypical statistical operator, and }$2$\textit{. The use of a escort
probability in terms of the heterotypical distribution of item 1.}

We reiterate that the \textit{escort probability }introduces corrections to
the insufficient description by including correlations and higher-order
variances of the observables involved. On the other hand, the \textit{%
heterotypical distribution} introduces corrections to the insufficient
description (or incomplete probabilities in Renyi's nomenclature) by
modifying the statistical weight of the dynamical states of the conventional
approach involved in the situation under consideration.

Moreover, we have considered a particular case, namely the statistics as
derived from the use of Renyi informational entropy. We centered the
attention on the derivation of an Unconventional Statistical Mechanics
appropriate for dealing with far-removed-from-equilibrium systems.

In conclusion, we may say that USM appears as a valuable approach, in which
the introduction of informational-entropic-indexes-dependent
informational-entro-pies leads to a particularly convenient and
sophisticated tool for fitting theory to experimental data for certain
classes of physical systems, for which the criterion of sufficiency in its
characterization cannot be properly satisfied. Among them we can pinpoint
fractal-like structured nanometric-scale systems, which, otherwise, would be
difficult to deal with within the framework of the conventional Statistical
Mechanics. While in the latter case one would need to have a detailed
description of the spatial characteristics of the structure of the system,
the other needs to pay the price of having an open adjustable index to be
fixed by best fitting with experimental results. It is relevant to notice
the fact that \textit{the infoentropic index(es) is(are) dependent on the
dynamics involved, the system's geometry and dimensions, boundary
conditions, its macroscopic-thermodynamic state (in equilibrium, or out of
it when becomes a function of time), and the experimental protocol}.

Finally, we call the attention to the fact that we have presented several
alternatives of infoentropies (see Table \textbf{I}), for which, as stated
in the main text, the uniform probability distribution is taken as the
reference one, and such generating functionals provide a corresponding
family of heterotypical probability distributions. However, other choices of
the reference probability can be made and then we have at our disposal
very-many possibilities: It is tempting to look for the construction of a
theory using for the probability of reference, instead of the uniform
distribution, Shannon-Jaynes informational-entropy in its incomplete
formalism, that is, when suffering from the deficiency that the researcher
cannot satisfy Fisher's criterion of sufficiency.\bigskip

{\large \textbf{ACKNOWLEDGMENTS}}\bigskip

We acknowledge financial support to our group provided in different
opportunities by the S\~{a}o Paulo State Research Foundation (FAPESP), the
Brazilian National Research Council (CNPq), the Ministry of Planning
(Finep), the Ministry of Education (CAPES), Unicamp Foundation (FAEP), IBM
Brasil, and the John Simon Guggenheim Memorial Foundation (New York, USA).

\newpage

\bibliographystyle{prsty}
\bibliography{bibliog}

\newpage

\begin{center}
\includegraphics[width=10cm]{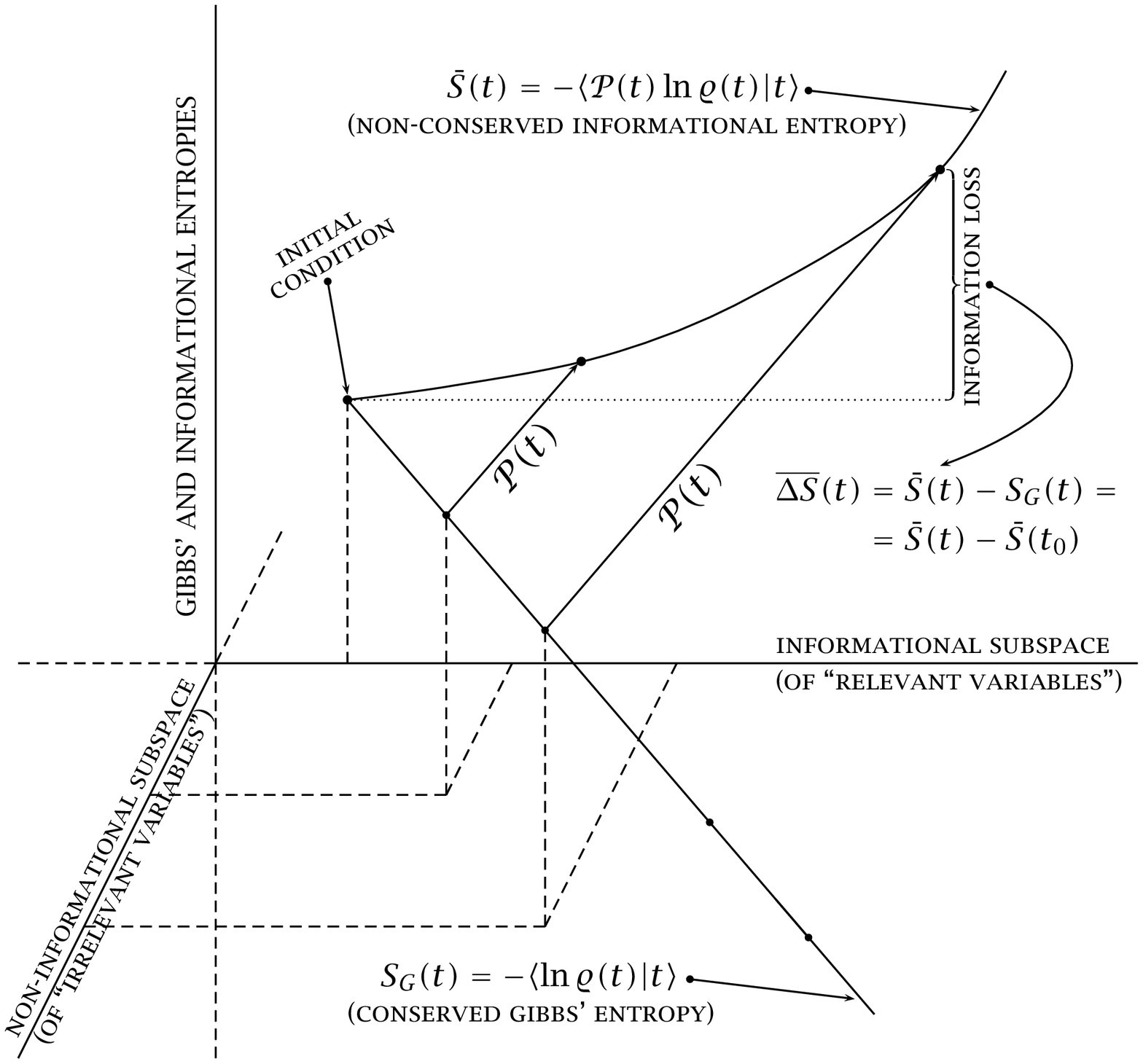}
\end{center}

\begin{description}
\item[Figure 1:]  An outline of the description of the
non-equilibrium-dissipative macroscopic state of the system. The projection
-- depending on the instantaneous macrostate of the system -- introduces the
coarse-graining procedure consisting into the projection onto the subspace
of the ``relevant'' variables associated to the informational constraints in
NESEF.
\end{description}

\end{document}